\documentclass[useAMS,usenatbib]{mn2e}
\usepackage{natbib}
\usepackage{graphicx}
\graphicspath{ {\images} }
\usepackage{savesym}
\usepackage{listings}
\usepackage{multirow}
\usepackage[intlimits]{amsmath}
\savesymbol{iint}
\usepackage{txfonts}
\usepackage{bm}
\usepackage{amssymb}
\usepackage{amsmath}
\usepackage{tabularx} 
\usepackage{color}
\usepackage{comment}
\usepackage{float}

\def\apj{Astrophys. J.}

\def\aa{Astron. Astrophys. }

\def\mnras{Mon. Not. Roy. Astron. Soc. }

\def\nat{Nature}

\newcommand{\be}{\begin{equation}}
\newcommand{\ee}{\end{equation}}
\newcommand{\bea}{\begin{eqnarray}}
\newcommand{\eea}{\end{eqnarray}}

\def\gsim{\mathrel{\raise.5ex\hbox{$>$}\mkern-14mu
             \lower0.6ex\hbox{$\sim$}}}
\def\lsim{\mathrel{\raise.3ex\hbox{$<$}\mkern-14mu
             \lower0.6ex\hbox{$\sim$}}}

\title[Pulsar PINN]{The Pulsar Magnetosphere with Machine Learning: Methodology}

\author[I. Contopoulos]
       {I. Dimitropoulos$^1$, I. Contopoulos$^2$\thanks{E-mail: icontop@academyofathens.gr}, V. Mpisketzis$^3$, E. Chaniadakis$^4$\\
$^1$ 
Department of Physics, University of Patras, Rio 26504, Greece\\
$^2$ 
Research Center for Astronomy and Applied Mathematics, Academy of Athens, Athens 11527, Greece\\
$^3$ 
Department of Physics, National and Kapodistrian University of Athens, Athens 15784, Greece\\
$^4$ 
Department of Informatics and Telecommunications, National and Kapodistrian University of Athens, Athens 15784, Greece
}

\begin{document}

\maketitle

\label{firstpage}

\begin{abstract}
{ 
In this study, we introduce a novel approach for deriving the solution of the ideal force-free steady-state pulsar magnetosphere in three dimensions. Our method involves partitioning the magnetosphere into the regions of closed and open field lines, and subsequently training two custom Physics Informed Neural Networks (PINNs) to generate the solution within each region. We periodically modify the shape of the boundary separating the two regions (the separatrix) to ensure pressure balance throughout. Our approach provides an effective way to handle mathematical contact discontinuities in Force-Free Electrodynamics (FFE). We present preliminary results in axisymmetry, which underscore the significant potential of our method. Finally, we discuss the challenges and limitations encountered while working with Neural Networks, thus providing valuable insights from our experience. 
}
\end{abstract}


\section{Introduction}

55 years after the discovery of pulsars (Hewish et al. 1968), there still does not exist a complete physical model for their strongly magnetized plasma-filled magnetospheres. The ground for their modelling was set by Goldreich \& Julian (1969), who suggested that one part of the magnetosphere rigidly corotates with the central star (the `closed line region' or `dead zone'), while the parts around the magnetic poles (the `polar caps') open up to infinity beyond the light cylinder. An electric field generated by the stellar rotation fills the magnetosphere with electric charges and the open field lines with electric currents. For 30 years, everyone’s focus was on the light cylinder, which was considered to be the source of magnetospheric dissipation. Our understanding changed dramatically when the first solution of the force-free ideal aligned pulsar magnetosphere was obtained (Contopoulos, Kazanas \& Fendt 1999; hereafter CKF). It was clearly proposed then that magnetospheric dissipation most probably takes place along the large scale magnetospheric electric current sheet that the new solution clearly identified, and not along the light cylinder. 

CKF introduced a novel iterative numerical method that relaxed to the steady-state solution of the axisymmetric pulsar magnetosphere. The non-axisymmetric problem, however, was solved with a different approach. Most studies of 3D pulsar magnetospheres are performed with time-dependent numerical simulations that start with a dipole magnetic field that is set into rotation at $t=0$. The simulations run for long enough times to attain a rotating steady-state configuration which is then presented as the solution of the pulsar magnetosphere. This has been the approach of the pioneering FFE (Force-Free Electrodynamics) simulations of Spitkovsky (2006) and Kalapotharakos \& Contopoulos (2009), the MHD (Magneto-Hydro-Dynamics) simulations of Tchekhovskoy, Spitkovsky \& Li (2013), the `ab initio' PIC (Particle-In-Cell) simulations of Philippov \& Spitkovsky (2014), Philippov, Spitkovsky \& Cerutti (2015), and Kalapotharakos et al. (2018), and most recently, the hybrid PIC-MHD simulations of Cerutti et al. (2022). These solutions have confirmed that the bulk of the magnetosphere operates under ideal force-free conditions, while electromagnetic dissipation, particle acceleration and high-energy radiation take place mostly along the equatorial current sheet. The steady-state problem was also solved with spectral methods (Parfrey, Beloborodov \& Hui 2012, P\'{e}tri 2012) which, unfortunately, behave very poorly in the equatorial current sheet. 

Time-dependent 3D numerical simulations have evolved dramatically in the past 10 years. Unfortunately, there exists no reference 3D ideal force-free solution with which to compare and validate their results. Thus, several important issues are left unanswered and new problems arose:
\begin{enumerate}
\item In all PIC simulations the magnetosphere opens-up a significant distance inside the light cylinder (their so-called `Y-point' may appear down to 85\% of the light cylinder radius; e.g. Hu \& Beloborodov 2022). If true, this would dramatically modify the pulsar spindown rate, the pulsar braking index, as well as the shape and spectrum of the high-energy pulsar radiation originating around the Y-point. We do not understand the physical origin of this effect in PIC simulations. We do not agree with the common explanation that it is due to the artificially high inertia of the PIC super-particles. Instead, we suspect that it is related to a subtle minimum of the magnetospheric electromagnetic energy (Contopoulos, Ntotsikas \& Gourgouliatos~2024). We still need to understand where the Y-point truly lies, and how fast it moves out as the pulsar spins down.


\item One major complication of pulsar magnetospheres is the appearance of electric current sheets inside which the ideal MHD approximation breaks down. Current sheets exist as mathematical contact-discontinuities across which $B^2-E^2$ is everywhere continuous (see below). MHD/FFE methods smooth out such discontinuities across several computational grid cells. Unfortunately, ideal FFE conditions are not valid inside current sheets. Moreover, different MHD numerical methods treat the equatorial current sheet differently (e.g. CKF and Timokhin~2006 erroneously place it inside the closed line region), and it is not possible in general to differentiate between artificial (numerical) and true (physical) features in the published solutions (e.g. interchanging multiple positive and negative charge layers as in Kalapotharakos et al.~2012, Mahlmann et al.~2022). Unfortunately, there does not exist a reference solution of the ideal (dissipation-less) force-free 3D problem against which to evaluate our numerical simulations. A reference solution exists only in 2D (CKF, Timokhin~2006).
        
    

\item Recent global PIC simulations (Cerutti, Philippov \& Dubus 2020, Hakobyan, Philippov \& Spitkovsky 2023) show that the electromagnetic (Poynting) flux remaining in the pulsar magnetosphere decreases as the logarithm of the distance r over the radius of the light cylinder $R_{\rm LC}$, namely $\propto 1-\beta_{\rm rec}\ln(r/R_{\rm LC})$. With $\beta_{\rm rec}\sim0.1$ as obtained from local PIC simulations of relativistic reconnection layers (e.g. Lyubarsky~2005, Werner et al.~2018), this results in the full pulsar magnetosphere completely disappearing via dissipation in the equatorial current sheet within a few hundred light cylinder radii. According to Hakobyan, Philippov \& Spitkovsky (2023), this logarithmic gradual electromagnetic energy dissipation applies to the full magnetosphere, not only to its undulating non-axisymmetric component as Cerutti, Philippov \& Dubus (2020) claim. This leaves no MHD pulsar wind to extend to the termination shock to power the X-ray pulsar nebula. We do not understand this result.
      

\item Time dependent simulations relax to one final steady-state solution. In nature, however, pulsars often exhibit mode-switching which cannot be accounted for by a unique solution. We must look for alternative global pulsar magnetospheric solutions, and we need to understand why and how pulsars may spontaneously jump between them (e.g. Ntotsikas et al.~2024).


\item The resolution of current PIC simulations is inadequate by several orders of magnitude to properly model the microphysics of the equatorial current sheet. In order to generate pulsar light curves and spectra that may be compared with observations, the simulation results (particle Lorentz factors, magnetic field values, electromagnetic spectra) are extrapolated by several orders of magnitude. Unfortunately, there is no agreement among different research groups on the particular method of extrapolation, and as a result, there is still no generally accepted understanding of the physical origin of the high-energy radiation from pulsars (e.g. synchrotron as in Hakobyan, Philippov \& Spitkovsky 2023; gamma-ray fundamental plane as in Kalapotharakos, Wadiasingh, Harding \& Kazanas 2022; inverse Compton as in Richards \& Lyutikov 2018; etc.). The remedy would be to investigate the trajectories of accelerated particles for realistic physical parameters (not PIC parameters) in the current sheet obtained in a reference solution.
\end{enumerate}
     
We believe that it is probably too early to derive safe conclusions about the pulsar magnetosphere from current global (`ab initio') PIC simulations, thus we can only trust them qualitatively (not quantitatively) to make meaningful comparisons with observations. Their parameters are orders of magnitude away from realistic values, and except for the aligned (non-pulsar) rotator, there is currently no way to validate their results by comparison with a reference solution. 
Given the inherent problems of all numerical methods discussed above, we believe that this is a dangerous development that may strongly mislead pulsar research.
We thus propose to return to the basics and independently obtain the reference ideal force-free magnetosphere with a novel numerical method. In the present paper we will present our first solutions of the axisymmetric problem, but our results will be directly generalized in the case of the 3D inclined rotator in a forthcoming publication. We present the general mathematical formulation of the problem in \S~2. In \S~3 we propose a novel numerical approach with machine learning that will allow us to obtain the solution in 3D. In \S~4 we formulate the problem in the axisymmetric case of the aligned rotator and present our first solutions of the pulsar equation with a trained Physics Informed Neural Network. 
{In the final section \S~5 we summarize our results and discuss
the challenges and limitations encountered while working with Neural Networks.
}

\section{General mathematical formulation}	

The magnetospheres of isolated neutron stars are generally considered to be dominated by the magnetic field. The physical conditions in the magnetosphere, i.e. the low mass of the charge carriers and the low density of the magnetospheric plasma, allow us to neglect gravity, thermal pressure and particle inertia as they are several orders of magnitude smaller than the electromagnetic forces. Particle inertia may only become important at the tip of the closed line region very close to the light cylinder. Other than that, force balance in the bulk of the pulsar magnetosphere is reduced to 
\begin{eqnarray}
&&\rho_{\rm e}{\bf E}+{\bf J}\times {\bf B}=0\ ,
\label{forcebalance}
\end{eqnarray}
where ${\bf B}$ and ${\bf E}$ are the magnetic and the electric field respectively, $\rho_{\rm e}$ is the electric charge density, and ${\bf J}$ is the electric current 
\begin{eqnarray}
{\bf J}
&\equiv& \frac{c}{4\pi}\nabla\cdot{\bf E}\ \frac{{\bf E}\times{\bf B}}{B^2}+ \frac{c}{4\pi}\frac{{\bf B}\cdot\nabla\times{\bf B}-{\bf E}\cdot\nabla\times{\bf E}}{B^2}\ {\bf B}\ .
\label{J}
\end{eqnarray}
The above general expression was obtained by Gruzinov~(1999) and Blandford (2002), and simplifies considerably in steady-state. 

All fields are calculated in the non-rotating inertial lab frame. In that frame, electromagnetic fields need to satisfy Maxwell's equations, namely
\begin{eqnarray}
&&\nabla\cdot{\bf B}=0\ ,\ 
\nabla\times{\bf B} = \frac{1}{c}\frac{\partial {\bf E}}{\partial t}+\frac{4\pi}{c}{\bf J}\ ,\nonumber\\
&& \nabla\cdot{\bf E}= \frac{4\pi}{c}\rho_{\rm e}\ ,\ 
\nabla\times{\bf E} = -\frac{1}{c}\frac{\partial {\bf B}}{\partial t}\ .
\end{eqnarray}

We will be searching for the steady-state solution to the above set of equations. Following the approach of Muslimov \& Harding~(2009), we will define the transformation of partial time derivatives between our lab frame, and a mathematical ({\it not physical}) reference frame with the pulsar angular velocity $\Omega$ around the axis $\theta=0$, namely
\begin{eqnarray}
\frac{\partial {\bf B}}{\partial t} & = & 
\left.\frac{\partial {\bf B}}{\partial t}\right|_{\rm corot}-
\nabla\times (r\sin\theta\ \Omega\ \hat{\phi}\times {\bf B})\ ,\nonumber\\
\frac{\partial {\bf E}}{\partial t} & = & 
\left.\frac{\partial {\bf E}}{\partial t}\right|_{\rm corot}-
\nabla\times (r\sin\theta\ \Omega\ \hat{\phi}\times {\bf E})
+r\sin\theta\ \Omega\ \hat{\phi}\ \nabla\cdot{\bf E}\ ,\nonumber\\
\frac{\partial \rho_{\rm e}}{\partial t} & = & 
\left.\frac{\partial \rho_{\rm e}}{\partial t}\right|_{\rm corot}
+r\sin\theta\ \Omega\ \hat{\phi}\cdot\nabla\rho_{\rm e}\ .
\end{eqnarray}
Time derivatives in that frame (denoted by the subscript `corot') vanish in steady state, and therefore, the Maxwell's equations that describe the steady state of the force-free pulsar magnetosphere become
\begin{eqnarray}
\nabla\times\left({\bf E}+\frac{r\sin\theta\ \Omega}{c}\hat{\phi}\times{\bf B}\right) &=& 0\ ,\label{E}\\
\nabla\times\left({\bf B}-\frac{r\sin\theta\ \Omega}{c}\hat{\phi}\times{\bf E}\right) &=& \frac{4\pi}{c}{\bf J}-\frac{r\sin\theta\ \Omega}{c}\hat{\phi}\ \nabla\cdot{\bf E}\ .\label{B}
\end{eqnarray}
Let us define here the light cylinder radius $R_{\rm LC}\equiv c/\Omega$. From eq.~(\ref{E}) above we obtain that
\begin{eqnarray}
&&{\bf E}\equiv {\bf E}_p=-\frac{r\sin\theta}{R_{\rm LC}}\hat{\phi}\times{\bf B}=
\frac{r\sin\theta}{R_{\rm LC}}(B_\theta,-B_r,0)
\label{E2}
\end{eqnarray}
in component form, where the subscript `p' denotes a poloidal component. Following eq.~(\ref{E2}), ${\bf E}\cdot \nabla\times {\bf E}=0$, and the expression for ${\bf J}$ in eq.~(\ref{J}) simplifies considerably. Substituting everything back in eq.~(\ref{B}) we obtain
$\nabla\times\{{\bf B}_p(1-(r\sin\theta/R_{\rm LC})^2)+{\bf B}_\phi\}\times {\bf B}=0$ or equivalently,
\begin{eqnarray}
&&
\nabla\times\left\{{\bf B}_p\left(1-\left(\frac{r\sin\theta}{R_{\rm LC}}\right)^2\right)+{\bf B}_\phi\right\}=\alpha {\bf B}\ .
\label{gpulsar}
\end{eqnarray}
Here, $\alpha$ is the force-free parameter that obeys the constraint
\begin{eqnarray}
&&{\bf B}\cdot\nabla\alpha=0\ ,
\label{alphaalongB}
\end{eqnarray}
namely that $\alpha$ is constant along magnetic field lines.
In other words, field lines lie along iso-contours of $\alpha$ in 3D. Notice that solving for the steady state configuration in the co-rotating frame has an important advantage over time-dependent simulations in the non-rotating (lab) frame. In the latter, the final configuration is rotating, i.e. time-dependent. { All its complex features (current sheets, Y-points, etc.) rotate in the simulation frame of reference, hence it is difficult to treat them and to guarantee that a steady-state is truly reached. On the other hand, solving for the steady-state configuration in the co-rotating frame is a relaxation approach that allows a better treatment of current sheets and more naturally reaches the final steady state.} Eq.~(\ref{gpulsar}) has first been obtained by Endean~(1974) and Mestel~(1975) but has never been solved before in 3D.

We now need to introduce the vector potential ${\bf A}$ such that ${\bf B}=\nabla\times{\bf A}$. We will work in spherical coordinates $(r,\theta,\phi)$ centered onto the central star.  Our notation will be slightly simplified if we introduce the vector magnetic flux components
\begin{eqnarray}
&&\Psi_r\equiv rA_r\ ,\ 
\Psi_\theta \equiv  rA_\theta\ ,\ 
\Psi_\phi \equiv  r\sin\theta\ A_\phi\ .
\label{fluxfunctions}
\end{eqnarray}
In that notation,
\begin{eqnarray}
B_r &\equiv& \frac{1}{r^2\sin\theta}\left(\frac{\partial \Psi_\phi}{\partial \theta}-\frac{\partial \Psi_\theta}{\partial \phi}\right)\ ,\nonumber\\
B_\theta &\equiv& \frac{1}{r^2\sin\theta}\left(\frac{\partial \Psi_r}{\partial \phi}-r\frac{\partial \Psi_\phi}{\partial r}\right)\ ,\nonumber\\
B_\phi &\equiv& \frac{1}{r}\frac{\partial \Psi_\theta}{\partial r}-\frac{1}{r^2}\frac{\partial \Psi_r}{\partial \theta}\ ,
\label{Bcomponents}
\end{eqnarray}
and eq.~(\ref{gpulsar}) becomes
\begin{eqnarray}
&&\frac{\partial^2\Psi_r}{\partial\phi^2}-r\frac{\partial^2\Psi_\phi}{\partial r\partial\phi}=\frac{r\sin\theta}{1-(r\sin\theta/R_{\rm LC})^2}
\left\{
-\alpha\left(\frac{\partial\Psi_\phi}{\partial\theta}-\frac{\partial\Psi_\theta}{\partial\phi}\right)\right.
\nonumber\\
&&+ \cos\theta\ \frac{\partial\Psi_\theta}{\partial r}-\frac{\cos\theta}{r}\frac{\partial\Psi_r}{\partial\theta}
+\left. \sin\theta\ \frac{\partial^2\Psi_\theta}{\partial\theta\partial r}-\frac{\sin\theta}{r}\frac{\partial^2\Psi_r}{\partial\theta^2}\right\}\ ,
\label{1}
\end{eqnarray}
\begin{eqnarray}
&&\frac{\partial^2\Psi_\theta}{\partial\phi^2}- r\frac{\partial^2\Psi_\phi}{\partial \theta\partial\phi}=
\frac{r\sin\theta}{1-(r\sin\theta/R_{\rm LC})^2}
\left\{
-\alpha\left(\frac{\partial\Psi_r}{\partial\phi}-r\frac{\partial\Psi_\phi}{\partial r}\right)\right.
\nonumber\\
&&
-\left. r\sin\theta\ \frac{\partial^2\Psi_\theta}{\partial r^2}
-\frac{\sin\theta}{r}\frac{\partial\Psi_r}{\partial\theta}+\sin\theta\ \frac{\partial^2\Psi_r}{\partial\theta\partial r}\right\}\ ,
\end{eqnarray}
\begin{eqnarray}
&&\frac{\partial^2\Psi_\phi}{\partial r^2}+\frac{1}{r^2}\frac{\partial^2\Psi_\phi}{\partial \theta^2}-
\frac{1}{r^2}\frac{\partial^2\Psi_\theta}{\partial\theta\partial\phi}
-\frac{1}{r}\frac{\partial^2\Psi_r}{\partial r\partial\phi}\nonumber\\
&&
+\frac{1}{r^2}\frac{\partial\Psi_r}{\partial\phi}
-
\frac{\cos\theta}{r^2 \sin\theta}\left(\frac{\partial\Psi_\phi}{\partial\theta}-
\frac{\partial\Psi_\theta}{\partial\phi}\right)=
\nonumber\\
&&
\frac{\sin\theta}{1-(r\sin\theta/R_{\rm LC})^2}\left\{
-\alpha\left(\frac{\partial\Psi_\theta}{\partial r}-\frac{1}{r}\frac{\partial\Psi_r}{\partial \theta}\right)\right.
\nonumber\\
&&
+\left.
\frac{2\cos\theta}{R^2_{\rm LC}}\left(\frac{\partial\Psi_\phi}{\partial\theta}
-\frac{\partial\Psi_\theta}{\partial\theta}\right)
-\frac{2\sin\theta}{R^2_{\rm LC}}\left(\frac{\partial\Psi_r}{\partial\phi}
-r\frac{\partial\Psi_\phi}{\partial r}\right)
\right\}
\label{3}
\end{eqnarray}
From the r.h.s. of eq.~(\ref{3}), the reader can check directly that the general regularization condition at the light cylinder $r\sin\theta=R_{\rm LC}$ becomes
\begin{equation}
\alpha=\left. 2\ \frac{\cos\theta\ B_r-\sin\theta\ B_\theta}{B_\phi R_{\rm LC}}\right|_{\rm LC}\equiv
\left.\frac{2B_z}{B_\phi R_{\rm LC}}\right|_{\rm LC}\ .
\label{regularization3D}
\end{equation}
Eq.~(\ref{regularization3D}) determines the value of the function $\alpha$ along all field lines that cross the light cylinder and never return to the star. In an untwisted magnetosphere, all other field lines that do not cross the light cylinder and form the closed line region have $\alpha=0$.

\section{Novel numerical approach}

We propose to solve the steady-state pulsar magnetosphere problem using Machine Learning techniques, and in particular by training a custom Neural Network (hereafter NN). In subsections 3.2 \& 3.3 below we will introduce two innovations that will lead to a better solution than what was obtained before both in 2D and 3D. Our formulation is general, but our method will also be implemented in axisymmetry in the next section.

\subsection{Machine learning}

A NN that is used to solve the partial differential equations that describe a physical problem is called a Physics Informed Neural Network (hereafter PINN). 
{ Quoting Wikipedia, "PINNs are a type of universal function approximators that can embed the knowledge of any physical laws that govern a given data-set in the learning process, and can be described by partial differential equations".} 
The entries of such NN are basically the spatial coordinates $r,\theta,\phi$, and the exits are $\Psi_r$, $\Psi_\theta$, $\Psi_\phi$ and $\alpha$. The NN consists of several internal layers (such a NN is termed a deep NN) with several tens of nodes each. Obviously, the exits of the NN are functions of the entries $r,\theta,\phi$, hence all order partial derivatives of the exits with respect to the entries are known. The fact that the derivatives of the fields are known locally is an important property of PINNs that differentiates them from finite difference grid methods.

{ The first implementation of PINNs in pulsar magnetospheres were obtained in the pioneering works of Stefanou et al. (2023, 2023b), where one can find a detailed description of their construction and training.
}
PINNs are trained with several loss functions that implement the constraints of the physical problem and are added together into a single loss function. The objective of the training is to adjust the internal parameters (weights) of the NN that will minimize the total loss calculated over random points in the computational domain, i.e. bring it as close to zero as possible everywhere. In other words, we choose random points over the computational domain $r_*\leq r\leq r_{\rm max}$, $0\leq \theta\leq \pi$, $0\leq \phi\leq 2\pi$ (here, $r_*$ is the radius of the central neutron star) and try to satisfy all the following constraints on them:
\begin{enumerate}
\item Eqs.~(\ref{1})-(\ref{3}) over the computational domain 
\item Eqs.~(\ref{alphaalongB}) over the computational domain 
\item Eq.~(\ref{regularization3D}) at $r\sin\theta =R_{\rm LC}$ (it may not be necessary to impose this condition since it is satisfied if (i) is satisfied)
\item $\alpha=0$ along field lines that do not cross the light cylinder (we will assume here an untwisted closed line region)
\item Boundary magnetic field conditions on the stellar surface. If we consider a central dipole magnetic field on a star of radius $r_*$ with the axis of the dipole along some direction $\hat{m}$ with a star-centered spherical coordinate system $(r,\theta_{m},\phi_{m})$ along that direction, then
\begin{eqnarray}
\Psi_r(r_*,\theta_{m},\phi_{m}) &=& 
\Psi_{\theta_{m}}(r_*,\theta_{m},\phi_{m}) = 0\nonumber\\
\Psi_{\phi_{m}}(r_*,\theta_{m},\phi_{m}) &=& \Psi_{\rm max}\sin^2\theta_{m}
\label{Psistar}
\end{eqnarray}
From the above star-centered components $\Psi_r,\Psi_{\theta_m},\Psi_{\phi_m}$ of the vector potential on the stellar surface, we will obtain the actual components $\Psi_r,\Psi_\theta,\Psi_\phi$ of the vector potential on the stellar surface as boundary conditions of our problem.
\end{enumerate}

One nice thing about NNs is that we can consider any number of conditions that the physical problem requires and, given enough internal layers and nodes, { training points and training steps,} the NN will manage to satisfy all of them by minimizing a global loss function { which is just the sum of the various constraints. Note that boundary conditions are also introduced in the form of extra constraints. For example, the loss functions that need to be minimized to zero and describe eq.~(\ref{Psistar}) are $|\Psi_r|, |\Psi_{\theta_m}|,$ and $|\Psi_{\phi_m}-\Psi_{\rm max}\sin^2\theta_m|$ summed over the $(r_*,\theta_m,\phi_m)$ training points chosen randomly along the stellar surface.} One other condition that needs to be satisfied everywhere is $E<B$. We have checked that this condition is automatically satisfied when magnetic field lines that cross the light cylinder open up to infinity, thus it is not necessary to impose it through a loss function term.

{ In our present PINN implementation, we used the PyTorch Deep Learning library. Code development was implemented in Python on Anaconda Jupyter Notebooks running on local GPUs, while production runs were performed in the Cloud in Google's Colab.}
What are the optimal number of internal nodes and training steps, the optimal activation functions and NN optimization parameters are not known a priori, and are determined after several trial trainings of the problem (see one particular implementation in \S~4.5 below). 
{ We implemented Adam optimizers, SiLU activation functions, and ReduceLROnPlateau schedulers from the PyTorch library.}
Another nice thing about NNs is that they are mesh-less, and the distribution of training points in the computational domain may be randomly chosen. 
This allows us to train the NN more precisely around the critical regions of interest (e.g. the separatrix between closed and open field lines, the Y-point at the tip of the closed line region, the equatorial current sheet, etc.). It also allows us to easily change the computational domain when the separatrix changes shape as we will see below. Both of the above benefits are very hard to implement in classical finite-difference numerical integration techniques with adaptive mesh refinement in a moving and deformable numerical grid. 
{ In our 2D axisymmetric runs (see \S~4 below) we used {1,000} random training points in the open field lines and {450} in the closed ones,  {200} random training points along the boundaries $r=r_*$ (the stellar surface), $q=0$ (infinity), $\theta=\pi/2$ (the equator), and {200} random training points along the separatrix. These random training points were updated every {5,000} training steps. We trained our PINNs for {50,000} training steps before we were satisfied with the loss convergence. One particular encouraging result was the reproduction of important known features of the solution, such as the distribution of the poloidal electric current in the open line region (see figure~\ref{figureresults2} below).
}


The general 3D PINN numerical procedure described above should relax to a solution where the closed line region extends up to some distance inside the light cylinder. The main problem with all numerical methods is that the solution of the pulsar magnetosphere contains electric current sheets along the separatrix and along the equator where the force-free ideal MHD condition breaks down. Unfortunately, the pulsar equation is not informed of their presence, and therefore, without special attention from the part of the programmer, it is incapable of supporting the mathematical contact discontinuities along them. This problem is addressed in the following subsection. One other problem is that, in past numerical methods, the imposition of the constraint $E<B$ was used to open up magnetic field lines that cross the light cylinder ($E<B$ is automatically satisfied inside the light cylinder where $r\sin\theta<R_{\rm LC}$). Therefore, if we want to study solutions where the closed line region does not extend up to the light cylinder, we must find other ways to open up the magnetosphere outside the closed line region. 
{One way may be to impose the constraint that $E<\nu B$ with $\nu<1$. 
In subsection~\ref{subsection3} below we propose yet another way to open up the magnetosphere. 
}

\subsection{Decomposition of the computational domain}

The first thing we want to do is to decompose the computational domain into the region of closed and open field lines. 
{ This is a central element of our methodology, namely that we choose from the very beginning of our simulation which field lines will be open and which ones will be closed. This is equivalent to an ad hoc determination from the very beginning of the extent of the polar cap (see below). In general, the solution that we will obtain will contain a closed line region that does not extend all the way to the light cylinder. It will, however, be a valid solution as we know since Contopoulos (2005) and Timokhin (2006). Up to this day it is not yet clear why in some time-dependent simulations the closed line region extends all the way to the light cylinder (e.g. Spitkovsky 2006, Kalapotharakos \& Contopoulos 2009, etc.), while in others it extends to only about $85\%$ of the light cylinder  (e.g. Hu \& Beloborodov 2022, Hakobyan et al. 2023). Which size of the polar cap corresponds to the physical (true) solution is a very important question that we want to elucidate in our next paper.}

\begin{figure}
 \centering
 \includegraphics[width=8cm,angle=0.0]{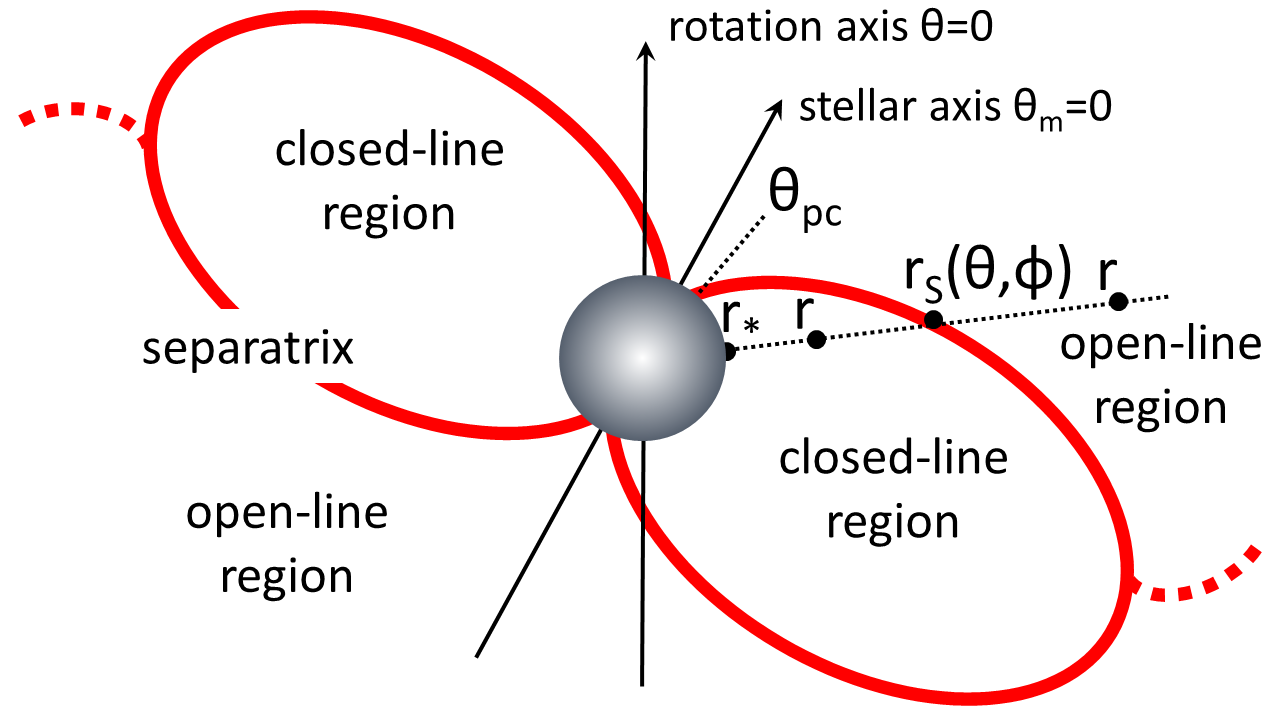}
 \caption{Schematic of the decomposition of the computational domain into a region of closed and open field lines. closed line region: between the central star $r=r_*$ and the separatrix $r=r_{\rm S}(\theta,\phi)$ (thick red line). open line region: outside the separatrix ($r>r_{\rm S}(\theta,\phi)$). 
In the open line region all field lines are artificially outflowing and there is no equatorial current sheet during the PINN training. When the PINN is trained, the open field lines in the southern hemisphere are reversed and the equatorial current sheet in the open line region appears along the surface of field-direction reversal (dotted red line).}
\label{figure}
\end{figure}

The two domains are separated by the separatrix surface characterized by the line where it originates on the surface of the star (see figure~\ref{figure}). In principle, one may obtain an analytic expression for the shape of the separatrix surface, namely
\begin{equation}
r_{\rm S}=r_{\rm S}(\theta,\phi)\ .
\label{rS}
\end{equation}
Obviously, the shape of the separatrix surface is unknown and must be determined. Let's assume for the moment that we choose ad hoc some initial form of $r_{\rm S}$ in eq.~(\ref{rS}) that originates around the magnetic poles of the central star at inclined polar angles $\theta_{m}=\theta_{\rm pc}$, $0\leq\phi_{m}\leq 2\pi$ and $\pi-\theta_{m}=\theta_{\rm pc}$, $0\leq\phi_{m}\leq 2\pi$. { Here, $\theta_{\rm pc}$ is the angular opening of the polar cap,} and $\theta_{m},\phi_{m}$ are spherical coordinates centered around the inclined magnetic axis $\hat{m}$ of the central star. 
Notice that for angles $\theta_{m}<\theta_{\rm pc}$ and $\theta_{m}>\pi-\theta_{\rm pc}$, the radius connecting the point $(r,\theta,\phi)$ with the origin does not cross the separatrix. 
%
Thus, in the open line region, the domain of integration becomes $r\geq r_*$, $0\leq \theta_{m}\leq \theta_{\rm pc}$ and $\pi-\theta_{\rm pc}\leq \theta_{m}\leq \pi$, and  $r\geq r_{\rm S}$, $\theta_{\rm pc}\leq \theta_{m}\leq \pi-\theta_{\rm pc}$.
In the closed line region, the domain of integration becomes $r_*\leq r\leq r_{\rm S}$, $\theta_{\rm pc}\leq \theta_{m}\leq \pi-\theta_{\rm pc}$.
{ The above $r,\theta_m,\phi_m$ intervals correspond to $r,\theta,\phi$ intervals and express mathematically what is obvious geometrically as closed line and open line regions in the schematic of figure~\ref{figure}.}
%

The reason we decided to separate the region of open and closed field lines is that, due to the electric current sheet that flows along the separatrix between the two regions, a strong contact discontinuity (of practically infinitesimal thickness) in the magnetic and electric fields exists along the separatrix. Uzdensky~(2003) (see also Lyubarskii~1990) integrated eq.~(\ref{forcebalance}) across current sheets and obtained that
\begin{equation}
(B^2-E^2)_{\rm below}=(B^2-E^2)_{\rm above}\ .
\label{B2E2}
\end{equation}
Treating such discontinuities inside any computational domain is very problematic and any computation method would generate spurious Gibbs oscillations around the separatrix. This is seen in all previous solutions of the pulsar magnetosphere (e.g. Kalapotharakos et al.~2012, Hu \& Beloborodov~2022, Mahlmann et al.~2022). We propose here a better way to treat such discontinuities: 
\begin{enumerate}
\item Solve eqs.~(\ref{1})-(\ref{3}) in the two regions independently 
for an initial arbitrary choice of the separatrix surface between them. An informed but not necessary unique initial choice would be the dipole magnetic field surface that originates at $\theta_{m}=\theta_{\rm pc}$ and $\theta_{m}=\pi-\theta_{\rm pc}$ on the stellar surface that corresponds to
\begin{equation}
\Psi_{\rm dipole}\equiv\Psi_{\rm max}\frac{r_*}{r}\sin^2\theta_{m}\equiv \Psi_{\rm max}\sin^2\theta_{\rm pc}\ ,
\end{equation}
which yields
\begin{equation}
r_{\rm S} \equiv  r_* \frac{\sin^2\theta_{m}}{\sin^2\theta_{\rm pc}}\ \mbox{if}\ \theta_{\rm pc}\leq \theta_{m}\leq \pi-\theta_{\rm pc}
\label{rS2}
\end{equation}
Obviously, for such an ad hoc shape of the separatrix surface, the quantity $B^2-E^2$ will in general be discontinuous across that surface.
\item Iteratively adjust the shape of the separatrix surface $r=r_{\rm S}(\theta,\phi)$ at all points $(\theta,\phi)$ so as to satisfy the condition that $B^2-E^2$ is continuous at all points of the separatrix.

In other words, we choose an initial dipolar shape for the separatrix surface, train the two PINNs in the open and closed line regions for a number of steps,
calculate $B^2-E^2$ at each point of the separatrix in the adjacent closed and open line regions, and displace { that point} proportionally to the difference between the two corresponding values in the direction of the smaller value. 
{ We must acknowledge that there is a high probability that the shape of the separatrix $r_{\rm S}(\theta,\phi)$ may not be a single-valued function of $\theta$ and $\phi$ for highly inclined rotators. We will thus take particular care in 3D to displace each separatrix point along the direction {\it perpendicular} to its surface.}
After each re-adjustment of the separatrix surface continue the training of the PINN for a few more steps, and then repeat the re-adjustment. This procedure has never been tried before and, as we will see below, yields the final solution where $B^2-E^2$ is continuous everywhere across the separatrix.
\end{enumerate}

\subsection{The numerical `disappearance' of the equatorial current sheet}
\label{subsection3}

{ One second improvement to the solution method has to do with the numerical treatment of the equatorial current sheet that originates at the tip of the closed line region.
The reason there exists an equatorial current sheet is that magnetic field lines leave one pole of the star, open up to infinity, and return from infinity to the other pole of the star. In doing so, they also carry the same amount of poloidal electric current in each hemisphere from each pole of the star to infinity. These two electric currents return to the star through the equatorial current sheet. In other words, the equatorial current sheet is there to close the global poloidal electric current circuit (CKF). It is thus obvious that, if we artificially (numerically) invert the direction of the field lines that leave the star from the southern pole, the electric current direction in the southern hemisphere will be inverted, and therefore there will be no need to close the global poloidal electric current circuit through an equatorial current sheet. 
This configuration is clearly artificial (it is equivalent to a magnetic monopole), but it is mathematically and dynamically equivalent to the configuration that we are investigating in the open line region, only without the mathematical discontinuity of the equatorial current sheet! We are able to implement this trick because we are treating the open line region independently from the closed line region. This is the same trick assumed by Bogovalov~(1999) when he obtained the solution for the tilted split monopole. We here generalize his approach and show that it is also valid (and very helpful) in the numerical treatment of the open line region in the more general dipole magnetosphere.
We have thus found a way to make the equatorial current sheet discontinuity `numerically disappear'.}

A numerical implementation of this trick is to assume that, 
%
in the open line region outside the closed line region, the boundary conditions for the magnetic flux function on the stellar polar caps will be
\begin{eqnarray}
\Psi_{\phi_{m}}(r_*,\theta_{m},\phi_{m})&=&\Psi_{\rm max}\sin^2\theta_{m}\ \mbox{for}\ 0\leq\theta_{m}\leq \theta_{\rm pc}\ ,\nonumber\\
\Psi_{\phi_{m}}(r_*,\theta_{m},\phi_{m})&=&2\Psi_{\rm S}-\Psi_{\rm max}\sin^2\theta_{m}\nonumber\\
&=&\Psi_{\rm max}(2\sin^2\theta_{\rm pc}-\sin^2\theta_{m})\nonumber\\
&&\ \ \ \ \ \ \ \ \ \ \ \ \ \ \ \ \ \ \ \ \ 
\mbox{for}\ \pi-\theta_{\rm pc}\leq \theta_{m}\leq \pi\nonumber\\
\Psi_{r}(r_*,\theta_{m},\phi_{m})&=&0\ ,\nonumber\\
\Psi_{\theta_{m}}(r_*,\theta_{m},\phi_{m})&=&0\ .
\label{Psitrick}
\end{eqnarray}
{ What we have done here is to express the boundary conditions in terms of the vector magnetic flux components $\Psi_r,\Psi_{\theta_m},\Psi_{\phi_m}$ in the inclined spherical system of coordinates $(r,\theta_m,\phi_m)$ in which a pure magnetic dipole is described as $\Psi_r=\Psi_{\theta_m}=0$ and $\Psi_{\phi_m}=\Psi_{\rm max}\sin^2\theta_m$.}
Obviously, in the open line region on the surface of the star, $0\leq \Psi_{\phi_{m}}\leq 2\Psi_{\rm S}$, whereas in the closed line region on the surface of the star, $\Psi_{\rm S}\leq \Psi_{\phi_{m}}\leq \Psi_{\rm max}$ ($\Psi_{\rm S}\equiv \Psi_{\rm max}\sin^2\theta_{\rm pc}$). This mathematical trick simplifies tremendously the numerical treatment of the large-scale open line region. 
{ Notice that, with or without the artificial flux reversal in the southern hemisphere, the pressure balance condition along the separatrix remains the same. After the solution is obtained, the field will be reversed in the southern hemisphere, and the true magnetic field configuration will be obtained with an equatorial current sheet along the surface of flux direction reversal.}

There is an important advantage of this method over standard numerical methods that include the equatorial current sheet in their domain of numerical integration. As $\theta$ approaches the equatorial current sheet at $\theta=\theta_{\rm eq.c.s.}$ from above, the first term in the expression for the field component $B_r$ in eq.~(\ref{Bcomponents}), namely $\partial\Psi_\phi/\partial\theta/(r^2\sin\theta)$, reaches in general a non-zero value that is immediately reversed below the equatorial current sheet. In other words,
\begin{equation}
\frac{\Psi_\phi}{\partial\theta}(r,\theta\rightarrow \theta_{\rm eq.c.s.}^-)=-\frac{\Psi_\phi}{\partial\theta}(r,\theta\rightarrow \theta_{\rm eq.c.s.}^+)\neq 0\ .
\label{Breq}
\end{equation}
Unfortunately, from the symmetry of $\Psi_\phi$ and its derivatives above and below the equatorial current sheet, in a numerical scheme without artificial field reversal in the southern hemisphere, the latitudinal derivative $\partial\Psi_\phi/\partial\theta$ will be equal to zero in the middle of the current sheet, which is not true in general. The numerical trick that we propose in this subsection naturally allows for $\partial\Psi_\phi/\partial\theta(r,\theta\rightarrow \theta_{\rm eq.c.s.})\neq 0$, which, after the final field reversal, naturally satisfies eq.~(\ref{Breq}).	

Finally, by choosing the angular size $\theta_{\rm pc}$ of the polar cap, this approach allows as to choose the amount of magnetic flux $\Psi_{\rm S}\equiv \Psi_{\rm max}\sin^2\theta_{\rm pc}$ that will open to infinity.\footnote{Notice that we can even choose a more general non-circular polar cap by specifying a non-uniform polar cap angular distribution $\theta_{\rm pc}(\phi_{m})$, as e.g. in P\'{e}tri~(2018). This will be implemented when we will adjust the azimuthal shape of the polar cap so that the separatrix surface touches the light cylinder at all azimuthal angles $\phi$. Notice also that, because we assume no twisting in the closed line region, the angular sizes of the north and south polar caps are the same in azimuth, i.e. the shape of the south polar cap is equal to $\theta_{\rm pc\ south}(\phi_{m})=\pi-\theta_{\rm pc\ north}(\phi_{m}$). This effect is caused because field lines in the corotating closed line region start and end along the same star-centered meridional angle $\phi_{m}$ and at the same latitudinal distances from each magnetic pole. This is also true for the last closed field lines immediately interior to the separatrix surface.} Due to the artificial field reversal, this amount is `forced' to extend to infinity, since it has no way to return to the star in the open line region. As we acknowledged above, we checked that the condition $E<B$ is automatically satisfied in the open line region beyond the light cylinder. 

\section{First results in axisymmetry}

\subsection{Mathematical formulation}

Before we proceed to a solution of the full 3D problem in a forthcoming publication, we would like to consider first the simpler axisymmetric problem ($\partial/\partial\phi=0$) where a steady-state solution is known (CKF). In the case of the aligned rotator it is customary to use the simplified notation $\Psi_\phi\equiv \Psi$. In units where $R_{\rm LC}\equiv 1$, eqs.~(\ref{Bcomponents}) yield
\begin{eqnarray}
B_r &\equiv& \frac{1}{r^2\sin\theta}\frac{\partial \Psi}{\partial \theta}\ ,\nonumber\\
B_\theta &\equiv& -\frac{1}{r\sin\theta}\frac{\partial \Psi}{\partial r}\ ,\nonumber\\
B_\phi &\equiv& \frac{I(\Psi)}{r\sin\theta}\ ,
\label{fieldcomponents2D}
\end{eqnarray}
and eq.~(\ref{3}) becomes
\begin{eqnarray}
&&
\left(1-r^2 \sin^2\theta\right) \left[ \frac{\partial^2 \Psi}{\partial r^2}-\frac{\partial \Psi}{\partial\theta}\frac{\cos\theta}{r^2 \sin\theta}+\frac{1}{r^2}\frac{\partial^2 \Psi}{\partial\theta^2}\right]\nonumber
\end{eqnarray}
\begin{equation}
-2r\sin\theta\left[\frac{\partial\Psi}{\partial\theta}\frac{\cos\theta}{r}+\frac{\partial \Psi}{\partial r} \sin\theta\right]+II'(\Psi)=0
\label{pulsareq1}
\end{equation}
Eq.~(\ref{pulsareq1}) is the well known pulsar equation (Wagoner \& Schalermann 1982) in spherical coordinates. Here, $I(\Psi)= r\sin\theta B_\phi$, and  $I'(\Psi)\equiv {\rm d}I/{\rm d}\Psi= \alpha$. The condition that $I=I(\Psi)$ (eq.~\ref{alphaalongB}) may be rewritten as
\begin{equation}
\frac{\partial\Psi}{\partial r}\frac{\partial I}{\partial\theta}-
\frac{\partial\Psi}{\partial \theta}\frac{\partial I}{\partial r}=0
\label{IparallelPsi}
\end{equation}
$I(\Psi)\ne 0$ only in the open line part of the magnetosphere. We consider here an untwisted magnetosphere with $I(\Psi)=0$ in its closed line part. Our results may also be generalized in the case of a twisted magnetosphere (see Ntotsikas et al.~2024). In axisymmetry, the regularization condition at the light cylinder (eq.~\ref{regularization3D}) becomes
\begin{eqnarray}
II'&=&2(\cos\theta\ B_r-\sin\theta\ B_\theta)|_{\rm LC}\equiv 2B_z|_{\rm LC}\ .
\label{alphareg}
\end{eqnarray}
As before, eq.~(\ref{alphareg}) determines the value of the function $II'(\Psi)$ along all field lines that cross the light cylinder and never return to the star. In an untwisted magnetosphere, all other field lines that do not cross the light cylinder and form the closed line region have $II'(\Psi)=0$. 

Eq.~(\ref{pulsareq1}) was first solved by Contopoulos, Kazanas \& Fendt~(1999) with a standard elliptic relaxation procedure based on a finite-difference grid in cylindrical coordinates. We will now solve the pulsar equation with machine learning.

\subsection{Machine learning}

In the aligned rotator, we need to obtain the solution $\Psi(r,\theta)$ and $I(\Psi)$. It is helpful to reformulate the problem in new coordinates 
\begin{eqnarray}
q &\equiv & \frac{1}{r}\nonumber\\
\mu &\equiv & \cos\theta\nonumber
\end{eqnarray}
In those new variables, eq.~(\ref{pulsareq1}) may be rewritten as
\begin{eqnarray}
&& (q^2-(1-\mu^2))\left(q^2\frac{\partial^2\Psi}{\partial q^2}
+(1-\mu^2)\frac{\partial^2\Psi}{\partial\mu^2}\right)\nonumber\\
&&+2q^3\frac{\partial\Psi}{\partial q}+2(1-\mu^2)\mu \frac{\partial\Psi}{\partial \mu}
+II'=0
\label{equationqmu1}
\end{eqnarray}
{ In our rescaled radial variable $q$, the outer radial boundary $r\rightarrow\infty$ lies at $q=0$. We tried to impose various boundary conditions at $q=0$ (e.g. $B_\theta=0$, or $\Psi=0$, or $\Psi=\Psi_{\rm S}=\mbox{const.}$), and in all cases the resulting poloidal field configuration became asymptotically radial. From this we concluded that, as long as the outer radial boundary of our calculation lies at infinity, the inner magnetospheric solution does not depend on the particular boundary conditions that we impose there.} In order to help the solution satisfy the boundary conditions along the axis and the stellar surface, namely $\Psi(r,\theta=0,\pi)=I(r,\theta=0,\pi)=0$ and $\Psi(r_*,\theta)=\sin^2\theta\ \Psi_{\rm max}$, we define new functions $f,{\cal I}$ as the new exits of the NN such that
\begin{eqnarray}
\Psi &\equiv& (1-\mu^2)f\nonumber\\
I &\equiv & (1-\mu^2){\cal I}
\end{eqnarray}
In these new functions, the pulsar equation (eq.~\ref{equationqmu1}) may be rewritten as
\begin{eqnarray}
&& (q^2-(1-\mu^2))\left(q^2\frac{\partial^2 f}{\partial^2 q}+(1-\mu^2) 
\frac{\partial^2 f}{\partial\mu^2}\right)\nonumber\\
&&+2q^3 \frac{\partial f}{\partial q} - (4\mu^2+2(q^2-(1-\mu^2))f\nonumber\\
&&+(2\mu(1-\mu^2)-4\mu (q^2-(1-\mu^2))\frac{\partial f}{\partial \mu}+{\cal S} = 0\ ,
\label{equationqmu}
\end{eqnarray}
while the condition that $I=I(\Psi)$ (eq.~\ref{IparallelPsi}) may be rewritten as
\begin{equation}
\frac{\partial f}{\partial q}\left(-2\mu{\cal I}+(1-\mu^2)\frac{\partial{\cal I}}{\partial\mu}\right)-
\frac{\partial {\cal I}}{\partial q}\left(-2\mu f+(1-\mu^2) \frac{\partial f}{\partial\mu}\right)=0
\label{IparallelPsi2}
\end{equation}
The source term in eq.~(\ref{equationqmu}) is equal to
\begin{eqnarray}
{\cal S} & \equiv & {\cal I}\frac{{\rm d}I}{{\rm d}\Psi}= {\cal I} \frac{|\nabla I|}{|\nabla\Psi|}\cdot{\rm sgn}\left(\nabla I \cdot \nabla\Psi\right)\nonumber\\
&=& {\cal I}\frac{\sqrt{q^2(1-\mu^2)\left(\frac{\partial{\cal I}}{\partial q}\right)^2+\left(2\mu{\cal I}-(1-\mu^2)\frac{\partial{\cal I}}{\partial\mu}\right)^2}}{\sqrt{q^2(1-\mu^2)\left(\frac{\partial f}{\partial q}\right)^2+\left(2\mu f-(1-\mu^2)\frac{\partial f}{\partial\mu}\right)^2}}\nonumber\\
&&
\cdot
{\rm sgn}\left(q^2(1-\mu^2)\frac{\partial{\cal I}}{\partial q} \frac{\partial f}{\partial q}\right.\nonumber\\
&&\ \ \ \ \ \ +\left.\left(2\mu{\cal I}-(1-\mu^2)\frac{\partial{\cal I}}{\partial \mu}\right)\left(2\mu f-(1-\mu^2)\frac{\partial f}{\partial \mu}\right)\right)
\end{eqnarray}

In this reformulation of the problem, the entries of the PINN are the spatial coordinates $(q,\mu)$, and the exits are $(f,{\cal I})$. The loss functions must then satisfy the conditions
\begin{enumerate}
\item Eq.~(\ref{equationqmu})
\item Eq.~(\ref{IparallelPsi2})
\item $f=1$ on the stellar surface
\item ${\cal I}=0$ in the closed line region
\end{enumerate}
{Notice that there is no special provision along the light cylinder as expressed by eq.~(\ref{regularization3D}), since if eq.~(\ref{equationqmu}) is satisfied, condition~(\ref{regularization3D}) is also automatically satisfied. This is a welcome advantage of the PINN methodology.
}


\subsection{Domain decomposition}

\begin{figure}
 \centering
 \includegraphics[width=8cm,angle=0.0]{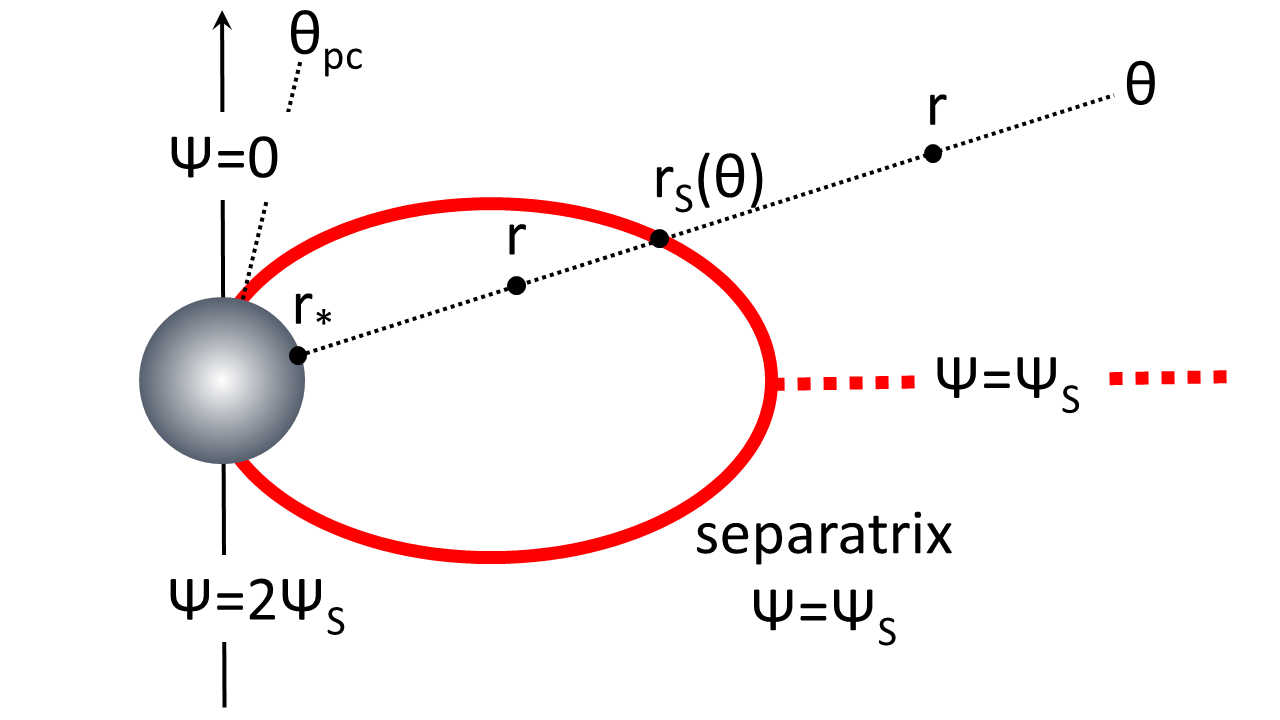}
\caption{Same as figure~\ref{figure} in axisymmetry. Note that we have implemented artificial field-direction reversal in the open line region.}
\label{figure2D}
\end{figure}

The domain decomposition is simpler in the case of the aligned pulsar where the direction $\hat{m}$  of the stellar magnetic dipole is along $\theta=0$, and therefore $\theta_{m}\equiv \theta$ and $\phi_{m}\equiv\phi$ (see figure~\ref{figure2D}). The open line region is $r\geq  r_{\rm S}$, $0\leq\theta\leq \pi$.
The closed line region is $r_*\leq r\leq r_{\rm S}$, $\theta_{\rm pc}\leq \theta\leq \pi-\theta_{\rm pc}$. Here, the initial choice of the separatrix radius $r_{\rm S}$ is
\begin{equation}
r_{\rm S}(\theta) = r_*\frac{\sin^2\theta}{\sin^2\theta_{\rm pc}}\ \ \ \ \ \mbox{if}\ \theta_{\rm pc}< \theta< \pi-\theta_{\rm pc}
\label{rS22D}
\end{equation}
The shape of $r_{\rm S}(\theta)$ will be adjusted iteratively in order to attain continuity of $B^2-E^2\equiv B_p^2(1-(r\sin\theta)^2)+B_\phi^2$, or equivalently continuity of
\begin{eqnarray}
&&(q^2-(1-\mu^2))\left(q^2 (1-\mu^2)\left(\frac{\partial f}{\partial q}\right)^2 + 4\mu^2 f^2 \right.\nonumber\\
&&+\left.  (1-\mu^2)^2 \left(\frac{\partial f}{\partial \mu}\right)^2 -4\mu(1-\mu^2)f  \frac{\partial f}{\partial \mu}\right)
+(1-\mu^2){\cal I}^2
\label{continuity}
\end{eqnarray}
across the separatrix as described in subsection~4.1 above. In the latter expression, ${\cal I}=0$ in the closed line region, and ${\cal I}\neq 0$ in the open line region across the separatrix.

We want to impose that $\Psi(r_{\rm S},\theta)=\Psi_{\rm S}=\Psi_{\rm max}\sin^2\theta_{\rm pc}$ in the separatrix between the two domains.
We, therefore, introduce one more loss function in both domains, namely one that requires
\begin{equation}
f\left(q=\frac{1}{r_{\rm S}},\mu\right)=\Psi_{\rm max}\frac{\sin^2\theta_{\rm pc}}{1-\mu^2}\ .
\end{equation}

\subsection{Field reversal}

The boundary conditions for the magnetic field in the closed line region $r_*\leq r\leq r_{\rm S}(\theta), \theta_{\rm pc}<\theta< \pi-\theta_{\rm pc}$ are
\begin{eqnarray}
\Psi(r_*,\theta)&=&\Psi_{\rm max}\sin^2\theta\ ,\nonumber\\
\Psi(r_{\rm S}(\theta),\theta)&=&\Psi_{\rm S}\ .
\end{eqnarray}
As before, $\theta_{\rm pc}$ is the opening of the polar cap, and $\Psi_{\rm S}\equiv \Psi_{\rm max}\sin^2\theta_{\rm pc}$. 
We implement the mathematical field reversal
in the open line region outside the closed line region with boundary conditions
\begin{eqnarray}
\Psi(r_*,\theta)&=&\Psi_{\rm max}\sin^2\theta\ \ \ \ \ \ \ \ \ \ \ \mbox{for}\ 0\leq\theta\leq \theta_{\rm pc}\ ,\nonumber\\
\Psi(r_*,\theta)&=&2\Psi_{\rm S}-\Psi_{\rm max}\sin^2\theta\nonumber\\
&=&\Psi_{\rm max}(2\sin^2\theta_{\rm pc}-\sin^2\theta)\ \ 
\mbox{for}\ \pi-\theta_{\rm pc}\leq \theta\leq \pi\ \ \nonumber\\
\Psi(r_{\rm S}(\theta),\theta)&=&\Psi_{\rm S}\ \ \ \ \ \ \ \ \ \ \ \ \ \ \ \ \ \ \ \ \ \ \ \  \mbox{for}\ \theta_{\rm pc}<\theta< \pi-\theta_{\rm pc}\ .
\end{eqnarray}
Obviously, in the open line region, $0\leq \Psi\leq 2\Psi_{\rm S}$, whereas in the closed line region, $\Psi_{\rm S}\leq \Psi\leq \Psi_{\rm max}$. This mathematical trick simplifies tremendously the numerical treatment of the large-scale open line region. There is no problem with the closed line region because the two regions are in contact along $\Psi(r,\theta)=\Psi_{\rm S}$.
After the solution is obtained, the field will be reversed in the open line region where $\Psi_{\rm S}\leq \Psi\leq 2\Psi_{\rm S}$, and the true magnetic field configuration will be presented with the equatorial current sheet along $\Psi=\Psi_{\rm S}$.

As we discussed above, one extra benefit of this approach is that $B_r(r>1,\theta=\pi/2^-)$ is naturally allowed to be non-zero and equal to $-B_r(r>1,\theta=\pi/2^+)$ across the equatorial current sheet. However, without the mathematical field reversal that we propose in the open line region, due to symmetry, $B_r=(\partial\Psi/\partial\theta)/(r^2\sin\theta)=0$ along the equator $\theta=\pi/2$. One might think that in a typical MHD/FFE grid simulation, the equatorial field reversal takes place within one grid cell. Unfortunately, in all known solutions in the literature the equatorial current sheet extends over several grid cells in thickness, which is not correct. This unfortunate result has the following repercussion: all components of the magnetic field are zero in a region of finite thickness immediately outside the Y-point. As a result, the magnetic (and electric) field pressure from immediately inside the tip of the closed line region pushes and protrudes outwards, forming a true Y-point as seen in all MHD/FFE numerical simulations. As is shown in detail in Contopoulos, Ntotsikas \& Gourgouliatos~(2024) however, this is not true, and the Y-point is in fact a T-point as first proposed by Uzdensky~(2003). This is captured in our present PINN solution with a proper treatment of the separatrix and equatorial current sheet via domain decomposition and field direction reversal in the open line region (see below).

\subsection{First results}

{ We implemented three independent NNs with multiple internal (hidden) layers:
}
\begin{enumerate}
\item  One PINN with two entries $q,\mu$, three hidden layers with 64 nodes each, and two exits $f,{\cal I}$. This PINN solves the pulsar equation in the open line region outside the separatrix { and is the most complex among the three}. It encounters no problems at the light cylinder where the functional form $I|_{\rm LC}=I(\Psi|_{\rm LC})$ is determined. Its training requires tens of thousands of steps, mainly because the condition that $I=I(\Psi)$ everywhere is very slowly enforced. 
The types of the individual activation functions and the weights of the individual losses are carefully chosen so as to yield consistent and stable convergence of the training.
\item  One PINN with two entries $q,\mu$, three hidden layers with 64 nodes each, and one exit $f$. This PINN solves the pulsar equation in the closed line region inside the separatrix where ${\cal I}=0$. The training of this PINN is stable and fast.
\item  The most crucial part of our method is the readjustment of the shape of the separatrix. We choose random angles $\theta_{\rm pc}<\theta<\pi-\theta_{\rm pc}$ where we know the values of $B^2(r_{\rm S},\theta)-E^2(r_{\rm S},\theta)$ both inside the outside the separatrix from the two PINNS trained in the closed and open line regions respectively. In general, these two values are not equal, so the radius of the separatrix must be adjusted at those angles. We implemented the adjustment
\begin{equation}
r_{\rm S\ new} = r_{\rm S}+2\beta\ \left(\frac{r_{\rm S}-r_*}{1-r_*}\right)^2\ 
\frac{(B^2-E^2)_{r=r_{\rm S}^-} -
(B^2-E^2)_{r=r_{\rm S}^+}}{(B^2-E^2)_{r=r_{\rm S}^-} +
(B^2-E^2)_{r=r_{\rm S}^+}}
\label{newrS}
\end{equation}
at those angles $\theta$. In the above expression, the separatrix is displaced due to the pressure imbalance in the third term of the right hand side product. The first term is introduced to avoid moving the separatrix footpoint on the stellar surface, 
and $\beta$ is an adjustable positive parameter. 
After the new position of the separatrix is defined on these randomly chosen angles $\theta_{\rm pc}<\theta<\pi-\theta_{\rm pc}$, a third NN with one entry $\mu$, two hidden layers with 128 nodes each, and one exit $r_{\rm S}$ is trained to yield the shape of the separatrix at all angles $\theta$ required by the first two main PINNs.
{ We have assumed here that the separatrix shape $r_{\rm S}(\theta)$ is a single-valued function of $\theta$. This is indeed true in axisymmetry, but is not in general true in highly-inclined oblique rotators. In that case we will displace each point of the separatrix in the direction perpendicular to its surface, not in the radial direction as in eq.~(\ref{newrS}).}
\end{enumerate}

{ The number of internal nodes and layers in the two PINNs that solve the pulsar equation was chosen by trial and error to be able to reproduce known features of the solution (e.g. the functional form of the electric current distribution $I=I(\Psi)$ shown in figure~\ref{figureresults2}). Our number of internal nodes per layer is {comparable} to that in the PINN implementation by Stefanou et al.~(2023b). The number of internal nodes in the third NN was chosen so that it describes in detail the deformed shape of the separatrix.
}

\begin{figure}
 \centering
 \includegraphics[width=9cm,angle=0.0]{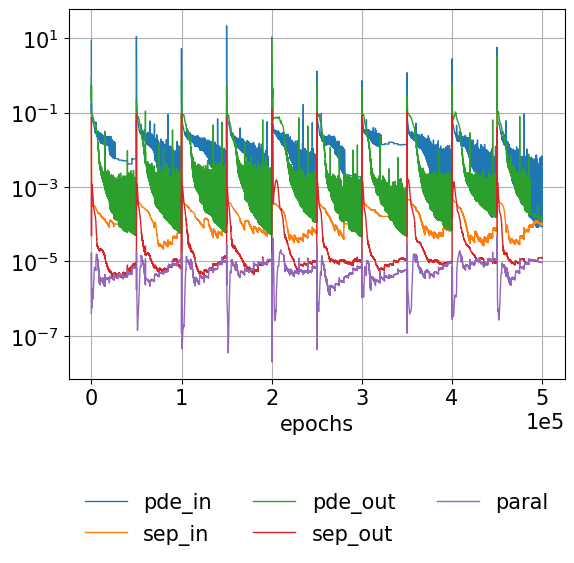}
\caption{ Evolution of the various losses (average square residuals) with the number of training steps for the two PINNs in the closed and open line regions. Initial PINN training: {50,000} steps. Subsequent separatrix shape adjustments and PINN retraining: $10\times$50,000~steps. Line notation for the closed/ open line regions respectively:
`pde~in'/ `pde~out': eq.~(\ref{equationqmu}), the pulsar equation (note that 
${\cal I}=0$ in the closed line region); 
`sep~in'/ `sep~out': separatrix boundary condition; `paral': eq.~(\ref{IparallelPsi2}),  $I=I(\Psi)$.  Abrupt loss changes correspond to updates in the training set.
}
\label{figureloss}
\end{figure}

{ 
In figure~\ref{figureloss} we show the evolution of our training losses as a function of the number of training steps. The various losses that we implemented to be minimized to zero are the averages of $(f-1)^2$ over the stellar surface, of $(\Psi-\Psi_{\rm S})^2$ along the separatrix and the equator in the open line region, and of course the average squares of the residuals of the central PDEs of the problem, eqs.~(\ref{equationqmu}) and (\ref{IparallelPsi2}).
The total loss is the sum of the above losses, {each one weighted by a corresponding adjustable weight factor (see discussion in \S~5). As we can see from figure~\ref{figureloss}, PDE errors (i.e. the residuals of eq.~\ref{equationqmu}) at the end of the training in the open line region are around $10^{-2}$, which, for magnetic flux values on the order of $\Psi\sim \Psi_{\rm max}=1$ and $(r_*/R_{\rm LC})\Psi_{\rm max}=0.25$ respectively, correspond to a conservative accuracy of about two significant digits in the value of $\Psi$. This is better than CKF, but worse than Timokhin~(2006) who mentions relative residuals on the order of $10^{-4}$, and can certainly be improved in the future with further development of our method.}

In practice, the two main PINNs are initially trained for {50,000} steps after which, satisfactory convergence is achieved in the two regions (all residual losses fall below $10^{-2}$).
After that initial training stage, the separatrix is displaced based on the resulting pressure differences between the closed and open line regions, and then the third NN is trained. The two main PINNs are re-trained for another {50,000} steps, and the process is repeated {10} times. In total, we run {500,000} training steps for the two main PINNs. 
}
In the resulting configuration, pressure balance is achieved across the separatrix to less than 1\%. 
{We further tested the accuracy of our PINN method in the Appendix by applying it to reproduce the standard vacuum magnetostatic dipole whose solution we know analytically ($\Psi(r,\theta)=\Psi_{\rm max}\sin^2\theta\ r_*/r$; figures~\ref{dipole} \& \ref{dipole2}). We can see there that the agreement with the analytic solution is {acceptable, at least in the region of interest inside the light cylinder where the density of training points is higher}.
}

\begin{figure}
 \centering
\hspace{0.3cm}
\includegraphics[width=8.5cm,,angle=0.0]{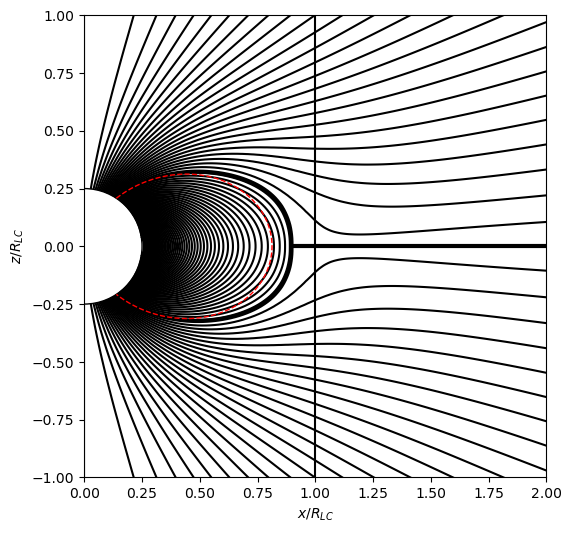}
 \caption{ Representative axisymmetric solution with domain decomposition and field reversal in the southern hemisphere.  Solution for $\theta_{\rm pc}=1.176 (r_*/R_{\rm LC})^{1/2}$, 
{or equivalently, $\Psi_{\rm open}\equiv \Psi_{\rm S}=1.23\Psi_{\rm dipole\ LC}$.}
Thin lines: poloidal magnetic field lines ($\Psi$ iso-contours {along integer multiples of $0.05\Psi_{\rm dipole\ LC}$}, with $\Psi=0$ along the $z$-axis). {Thin red dotted line: initial dipole position of separatrix surface.} Thick lines: separatrix $\Psi_{\rm open}\equiv \Psi_{\rm S}=\Psi_{\rm max}\sin^2\theta_{\rm pc}$ and equatorial current sheet. 
Vertical line: light cylinder. 
Notice the shape of the tip of the closed line region: in agreement with Uzdensky~2003, its true shape is a T-point. }
\label{figureresults}
\end{figure}
\begin{figure}
\includegraphics[width=6.8cm,angle=0.0]{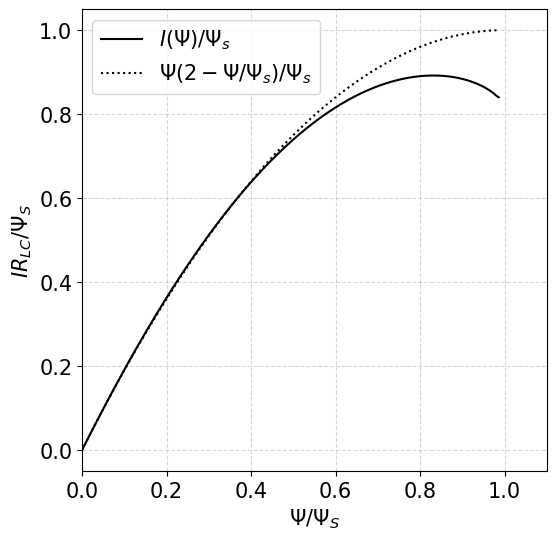}
\includegraphics[width=6.8cm,angle=0.0]{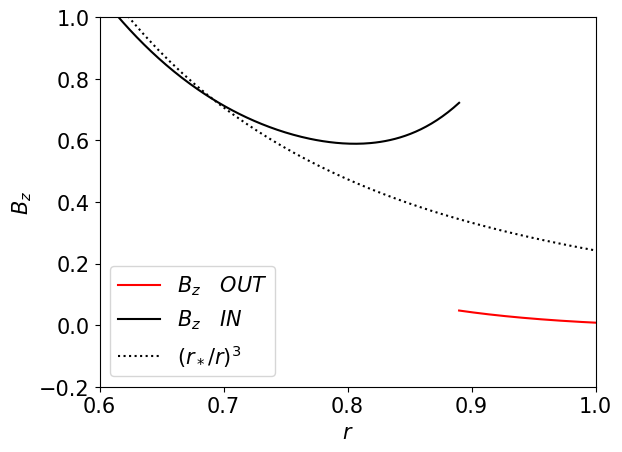}
\includegraphics[width=6.8cm,angle=0.0]{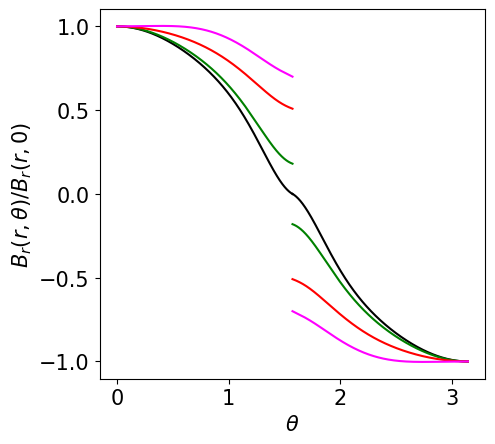}
 \caption{Top: electric current distribution $I(\Psi)R_{\rm LC}/\Psi_{\rm S}$ across open magnetic field lines $0\leq \Psi/\Psi_{\rm S}\leq 1$ for the solution of figure~\ref{figureresults}. 
Dotted line: analytic expression for split monopole.
{
Our result is in excellent agreement with figure~3 of Timokhin~2006. We consider this an important success of the PINN methodology.
Middle: the distribution of $B_z(r,\theta=\pi/2)$ inside (black line) and outside (red line) the Y-point. The dotted line represents the standard $1/r^3$ dipolar dependence. We clearly infer both the outward stretching of the initially dipolar field (at small $r$ distances the solid line lies below the dotted one), and the inwardx squeezing of the field right behind the Y-point that is required to satisfy pressure balance (in agreement with figure~11 of Timokhin~2006).
Bottom: the distribution of $B_r(r,\theta)/B_r(r,0)$ for $r=R_{\rm Y}$/$R_{\rm LC}$/$2R_{\rm LC}$/$3R_{\rm LC}$ (black/green/red/magenta lines respectively). We see the gradual transition from a dipole field ($B_r(r,\theta)=B_r(r,0)\cos\theta$) to a split-monopole field ($B_r(r,\theta)=B_r(r,0){\rm sgn}(\cos\theta)$) with distance from the star.
}
Notice that $B_r(r,\theta=\pi/2^-)=-B_r(r,\theta=\pi/2^+)\neq 0$ (without field reversal, $B_r(r,\theta=\pi/2)$ would erroneously be equal to zero).}
\label{figureresults2}
\end{figure}

{ We present our preliminary results in figures~\ref{figureresults} and \ref{figureresults2} for $\theta_{\rm pc}=1.176(r_*/R_{\rm LC})^{1/2}$ and $\Psi_{\rm open}\equiv \Psi_{\rm S}=\Psi_{\rm max}\sin^2\theta_{\rm pc}$ for $r_*=0.25R_{\rm LC}$. This particular value of $\theta_{\rm pc}$ was specially chosen with the following in mind: It is straightforward to calculate that, in a dipolar magnetic field configuration, the magnetic field line that crosses the light cylinder corresponds to $\Psi_{\rm dipole\ LC}=\Psi_{\rm max}r_*/R_{\rm LC}$, and $\theta_{\rm dipole\ pc}={\rm arcsin}[ (r_*/R_{\rm LC})^{1/2}]\approx  (r_*/R_{\rm LC})^{1/2}$. In previous high-resolution solutions of the pulsar equation (e.g. Timokhin~2006), the magnetic field line that reaches the light cylinder corresponds to $\Psi_{\rm LC\ Timokhin}=1.23 \Psi_{\rm dipole\ LC}$ and $\theta_{\rm pc\ Timokhin}={\rm arcsin}[ (1.23r_*/R_{\rm LC})^{1/2}]= 1.176(r_*/R_{\rm LC})^{1/2}$ for $r_*=0.25R_{\rm LC}$. Therefore, this particular value of $\theta_{\rm pc}$ (and correspondingly $\Psi_{\rm open}$) was chosen because in the standard solution (e.g. Timkokhin 2006) the Y-point lies very close to the light cylinder.

{We found that $R_{\rm Y}= 0.88 R_{\rm LC}$ which is outside  $R_{\rm dipole}(\Psi_{\rm S})=r_*/\sin^2\theta_{\rm pc}= 0.81R_{\rm LC}$, as expected from previous solutions, but also inside $R_{\rm LC}$.  We suspect that this is due to our treatment of the separatrix which differs from that in all previous MHD/FFE/PIC solutions. In the present work, the separatrix is treated as a perfect (zero thickness) mathematical discontinuity, and is found to lie at a position where pressure balance is satisfied as expressed by eq.~(\ref{B2E2}). In all previous solutions, however, the separatrix return current is distributed among a bundle of magnetic field lines that contains a finite amount of poloidal magnetic flux $\delta\Psi\neq 0$ (e.g.  $\delta\Psi=0.01-0.06 \Psi_{\rm S}$  in Timokhin~2006). The outer part of this bundle represents the last closed field line that extends very close to the light cylinder and reaches the equator at a non-vertical angle. This position is what is usually characterized as the Y-point. The separatrix return current, however, is distributed till the inner part of this bundle which represents the last closed field line without poloidal electric current. This reaches the equator vertically inside the light cylinder (see figure~4 in Timokhin~2006, figures 2 \& 3 in Contopoulos et al.~2024 for details; see also figure~1 in Hakobyan et al.~2023). In other words, in all previous numerical solutions there is no well-defined Y-point, only an extended region of distributed separatrix return current. We do not know whether the separatrix return current is indeed distributed. This merits further investigation which is, however, outside the scope of the present work. Under our present assumption that the separatrix return current is not distributed  over $\Psi$, $R_{\rm Y}(\Psi_{\rm S})$ is found to lie slightly inside the light cylinder.
}

}

Another interesting point is the squeezing of the separatrix surface in the latitudinal direction right above the stellar surface. This is due to the pressure of the nonzero toroidal field $B_\phi^2$ in the open line region. Furthermore, as expected, the Y-point is indeed a T-point, i.e. the separatrix crosses the equator vertically. In fact, the closed line region is squeezed inwards at its tip on the equator, which results in the slight increase of $B_z(r=R^-_{\rm Y},\theta=\pi/2)$ that is required to satisfy pressure balance accross the T-point (the superscript $-$ is used to denote `right inside the Y-point'; figure~\ref{figureresults2} middle; see also figure~11 in Timokhin~2006). Moreover, $B_r(r>R_{\rm Y},\theta=\pi/2^-)=-B_r(r>R_{\rm Y},\theta=\pi/2^+)\neq 0$ (superscripts $-/+$ are used to denote `right above/below the equator' respectively; figure~\ref{figureresults2} bottom). The above observations merit further investigation in a future publication.


\section{Discussion and conclusions}

In this work, we have presented a novel numerical method for solving the generalized pulsar equation with PINNs. We have observed that PINNs behave satisfactorily around the light cylinder, but fail to properly treat the separatrix current sheet. For this reason, we have also introduced two innovations, namely the decomposition of the computation domain into two regions (one inside the separatrix current sheet and one outside), and the mathematical field reversal in the open line region. The latter allows us to completely avoid the equatorial current sheet discontinuity outside the light cylinder. The former allows PINNs to relax to a continuous and smooth solution that does not contain any mathematical singularities. The domain decomposition and field reversal that we introduced may also be implemented in more general situations with MHD current sheets, as for example in active regions of the solar corona. Regarding computational cost, axisymmetric PINN training is completed in a few hours on one GPU, whereas standard high-resolution solutions of the pulsar equation require about 1 day on a standard PC. Standard HPC time-dependent 3D simulations require days, and we can only hope that extrapolation of our method to an oblique rotator will be faster. 
{
Although the PINN methodology has better behavior around the light cylinder and slightly faster convergence, we do not see any significant advantage over more traditional methods. One could also use standard numerical techniques to treat the deformable regions of open and closed field lines by constructing curvilinear grids that adapt to the evolving geometry of the boundary between the two regions.
}

{While NNs possess considerable power, it’s important to recognize their inherent challenges and limitations. Without careful supervision, there’s no assurance that they will converge to the accurate solution, even after hundreds of thousands of iterations.
Quoting from Moschou et al. 2023, {\it ``PINNs suffer from competing losses during gradient descent that can lead to poor performance especially in physical setups involving multiple scales''}. Our experience has shown that NNs can easily become ‘trapped’ in a configuration with substantial losses, in particular when the set of training points is not balanced, i.e. if one loss is trained over many more training points than others. As such, users of NNs must be vigilant, particularly in increasing the weights of those losses that appear to be inactive during their training. It is crucial that the training process is never left unmonitored. Users should track the evolution of the losses and may need to calibrate their weights. With sufficient experience working with NNs, users can develop an understanding of when the training process is failing and when a loss weight readjustment is necessary. This hands-on approach ensures the effective application of NNs and mitigates potential issues. 
}


%
%

{ We have obtained a preliminary solution of the pulsar equation for a particular value of the open magnetic flux $\Psi_{\rm open}=\Psi_{\rm max}\sin^2\theta_{\rm pc}$ with $\theta_{\rm pc}=1.176(r_*/R_{\rm LC})^{1/2}$,
in which the closed line region is found to extend radially up to $0.88R_{\rm LC}$.
} We would like to correct here an important misunderstanding found in the literature. With the solution obtained in \S~4 above, we confirm Uzdensky~(2003) that in the presence of the separatrix return current sheet, the Y-point is in fact a T-point at its tip. This means that at its tip it has a finite height and it is not at a non-vertical angle with respect to the equator as Gruzinov~(2005) surmises. This realization makes a huge difference in the calculation of the electromagnetic field energy contained in this region of finite height at the tip of the closed line region which diverges as the Y-point approaches the light cylinder (see Contopoulos, Ntotsikas \& Gourgouliatos~2024 for a detailed analysis of this effect). This result differs from Gruzinov~(2005) that the magnetospheric energy contained around the Y-point is finite (that result was based on his conclusion that the separatrix arrives on the equator at a non-vertical angle; in that case, the electromagnetic energy contained in the tip of the closed line region indeed remains finite). 
This could be another reason why the Y-point is located well inside the light cylinder in all high-resolution global PIC simulations conducted in the last decade.

In summary, our  implementation differs from Stefanou et al.~(2023b) who were the first to solve the pulsar equation with a PINN and their work was an inspiration for our present work. We were the first to introduce a proper treatment of mathematical contact discontinuities in FFE.
This yielded important details that differ from previous solutions of the pulsar magnetosphere (e.g. the shape and the extent of the tip of the closed line region). A more detailed description of our results in the axisymmetric magnetosphere will be presented in a follow-up publication. A detailed solution of the 3D magnetosphere where the real power of our method will be manifested will also be presented in the near future.

\section*{Acknowledgements}

We would like to thank Petros Stefanou and Jose Pons for interesting and inspiring discussions. {We also thank the anonymous referee for insisting that we should improve the accuracy of our preliminary results. This led us to acknowledge the challenges and limitations of NNs.} We would also like to thank the International Space Science Institute (ISSI) for providing financial support for the organization of the meeting of ISSI Team No 459 led by I. Contopoulos and D. Kazanas where the ideas presented in the paper originated. { This research work was supported by the Hellenic Foundation for Research and Innovation (HFRI) under the $4^{\rm th}$ Call for HFRI PhD Fellowships (Fellowship Number: 9239).}

\section*{Data availability statement}
The data underlying this article will be shared on reasonable request to the corresponding author.


\bibliographystyle{mn2e}

{
\section*{Appendix}

In order to test the accuracy of our method, we applied it to solve the simple magnetostatic problem $\nabla\times{\bf B}=0$, or equivalently
\begin{eqnarray}
&& q^2\frac{\partial^2 f}{\partial^2 q}+2q \frac{\partial f}{\partial q}
 - 2f-4\mu \frac{\partial f}{\partial \mu}+(1-\mu^2) 
\frac{\partial^2 f}{\partial\mu^2} = 0\ 
\label{equationdipole}
\end{eqnarray}
in our notation, with boundary condition $\Psi(r_*,\theta)=\Psi_{\rm max}\sin^2\theta$ on the stellar surface. In this problem there is no rotation, thus there is no poloidal electric current $I$. Nevertheless, we used the formal methodology of our present paper, namely domain decomposition and separatrix re-adjustement. 
{ 
It is also implied that we used a denser distribution of training points in the region of interest of the rotating problem inside the light cylinder.
}
Notice that we chose the value $\Psi_{\rm S}=1.23\Psi_{\rm dipole\ LC}$ ad hoc since in this problem there is no light cylinder, thus there is no distinction between open and closed field lines as in the real pulsar problem. The initial separatrix position was taken to be non-dipolar in order to allow the method to obtain its final position (thick black line).
In figure~6 we plot the outcome of the PINN $\Psi\equiv (1-\mu^2)f$ (solid lines) and compare it with the exact solution $\Psi_{\rm dipole}=\Psi_{\rm max}\sin^2\theta\ r_*/r$ (dashed lines).
{We overplot in color the accuracy of the solution defined as $|(\Psi-\Psi_{\rm dipole})|/\Psi_{\rm dipole}$. We see that the accuracy of our method is better than $1.5\%$ in the largest part of the magnetosphere.}
The two solutions are practically indistinguishable inside the distance that corresponds to the light cylinder of the rotating solution
{where the density of training points is higher (accuracy around $0.5\%$).
}
{
Notice that in figure~6 we tested the accuracy of both the outcome of the PINN and the movement of the separatrix. In order to test the accuracy of the PINN itself we solve eq.~(\ref{equationdipole}) without a separatrix. This yields the solution shown in figure~7 which has even higher accuracy (better than $0.2\%$).
}

{
We conclude that the numerical accuracy of our method is not on par yet with that of standard methods such as finite difference, finite volume or spectral methods. This is also the case in the PINN solution of Stefanou et al.~(2023b). On the other hand, the numerical treatment of current sheets by standard methods it too is problematic (numerical current sheets have unphysical thicknesses  and the numerical dissipation there is questionable). The purpose of this study has been the determination of the location of the current sheet and the imposition of pressure balance along that surface. This has never been successfully attained before in 3D, thus it is worthwhile to apply our methodology to the oblique rotator too. In order to improve our results in future iterations of this work, we may try to combine standard methods outside current sheets with our methodology of moving the separatrix. This is still work in progress. 
}

\begin{figure}
 \centering
\includegraphics[width=8.5cm,angle=0.0]{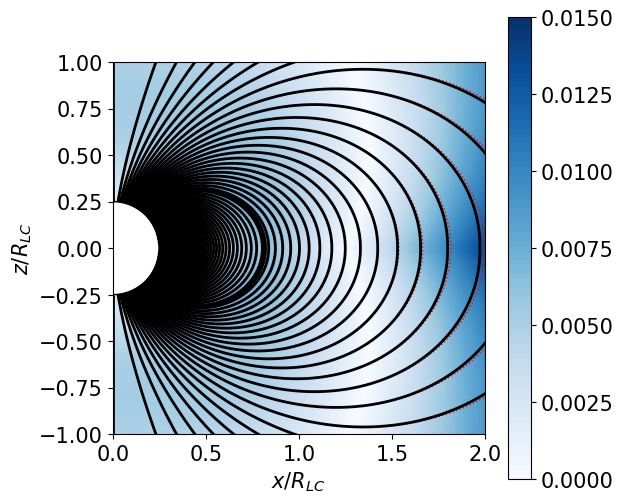}
 \caption{{Solution to test the PINN methodology together with separatrix adjustment. Hyperparameters as in \S~4.5.} $\Psi$ and separatrix  iso-contours as in figure~4. Red dotted lines: magnetostatic dipole with known analytic solution. 
{Color map: accuracy defined as $|(\Psi-\Psi_{\rm dipole})|/\Psi_{\rm dipole}$}.
In order to test the convergence of the method, we initialized the surface $\Psi=\Psi_{\rm S}=1.23\Psi_{\rm dipole\ LC}$ at a position different from its final dipole position. 
{Accuracy is better than $1.5\%$ in the largest part of the magnetosphere and around $0.5\%$ inside the distance that corresponds to the light cylinder of the rotating solution where the density of training points is higher.
}
}
\label{dipole}
\end{figure}

\begin{figure}
 \centering
\includegraphics[width=8.5cm,angle=0.0]{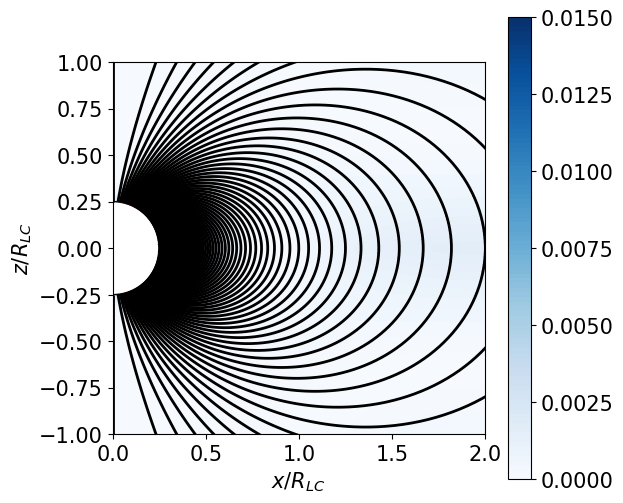}
 \caption{Solution to test the PINN methodology alone (no separatrix adjustment). Plot parameters as in figure~6. The numerical solution (solid lines) is indistinguishable from the analytic solution (red dotted lines). In the largest part of the magnetosphere the accuracy is well below $0.2\%$.
}
\label{dipole2}
\end{figure}

\end{document}